\def\a{\alpha}\def\b{\beta}\def\d{\delta}
\def\h{\theta}
\def\k{\kappa}\def\l{\lambda}\def\m{\mu}\def\n{\nu}\def\o{\omega}\def\r{\rho}\def\s{\sigma}

\def\D{\Delta}
\def\O{\Omega}

\def\ha{{1\over 2}}
\def\qu{{1\over 4}}

\def\({\left(}\def\){\right)}\def\[{\left[}\def\]{\right]}
\def\lra{\leftrightarrow}

\def\mn{{\mu\nu}}\def\ij{{ij}}

\def\coo{coordinates }

\def\dof{degrees of freedom }

\def\poi{Poincar\'e }

\def\wrt{with respect to }\def\ie{i.e.\ }

\def\cor{commutation relations }

\def\nc{noncommutative }\def\kp{$\k$-\poi }

\def\section#1{\bigskip\noindent{\bf#1}\smallskip}

\def\subsection#1{\smallskip\noindent{\it#1}\smallskip}
\def\nota{\footnote{$^\dagger$}}

\def\acknowledgments{\bigskip\centerline{\bf Acknowledgments}\smallskip\noindent}

\def\PL#1{Phys.\ Lett.\ {\bf#1}}
\def\PRL#1{Phys.\ Rev.\ Lett.\ {\bf#1}}
\def\PR#1{Phys.\ Rev.\ {\bf#1}}
\def\NP#1{Nucl.\ Phys.\ {\bf#1}}

\def\JHEP#1{JHEP\ {\bf#1}}

\def\ref#1{\medskip\everypar={\hangindent 2\parindent}#1}
\def\beginref{\begingroup
\bigskip
\centerline{\bf References}
\nobreak\noindent}
\def\endref{\par\endgroup}

\def\hx{\hat x}\def\hp{\hat p}

\def\rs{{\r\s}}

\def\eha{extended Heisenberg algebra }
\def\esa{extended Snyder algebra }\def\oa{object of noncommutativity }
\def\ad{a^\dagger}\def\nm{{\n\m}}

{\nopagenumbers
\line{}
\vskip40pt
\centerline{\bf Quantum mechanics of the extended Snyder model}
\vskip40pt
\centerline{{\bf S. Meljanac}\nota{e-mail: meljanac@irb.hr}}
\vskip5pt
\centerline {Rudjer Bo\v skovi\'c Institute, Theoretical Physics Division}
\centerline{Bljeni\v cka c. 54, 10002 Zagreb, Croatia}
\vskip10pt
\centerline{and}
\vskip5pt
\centerline{{\bf S. Mignemi}\nota{e-mail: smignemi@unica.it}}
\vskip5pt
\centerline {Dipartimento di Matematica, Universit\`a di Cagliari}
\centerline{via Ospedale 72, 09124 Cagliari, Italy}
\smallskip
\centerline{and INFN, Sezione di Cagliari}

\vskip40pt
{\noindent\centerline{\bf Abstract}
}
\vskip10pt
We investigate a quantum mechanical harmonic oscillator based on the extended Snyder model.
This realization of the Snyder model is constructed as a quantum phase space generated by $D$ spatial coordinates and $D(D-1)/2$ tensorial degrees of freedom,
together with their conjugate momenta. The \coo obey nontrivial \cor and generate a noncommutative geometry, which admits nicer properties than the usual
realization of the model, in particular giving rise to an associative star product.

The spectrum of the harmonic oscillator is studied through the introduction of creation and annihilation operators.
Some physical consequences of the introduction of the additional degrees of freedom are discussed.

\vfil\eject}

\section{1. Introduction}

The Snyder model [1] is known to be the first proposal of quantized spacetime. It is based on an algebra generated by spacetime coordinates
and Lorentz generators, that allows nontrivial \cor between position operators without breaking the Lorentz invariance.

Although in its time it did not attract much attention,
its relevance increased when new models, as Moyal space or \kp algebra [2,3], and methods related to noncommutative geometry [4] were introduced.
In particular, its formulation in terms of Hopf algebras was investigated in [5]. In that paper the coproduct and the star product
were calculated for the algebra generated by the noncommuting position operators.

However, in Snyder algebra the \cor of the position coordinates do not close, since they give rise to Lorentz generators,
and therefore the structure obtained in [5] is not strictly a Hopf algebra, since it is nonassociative.
A way to obtain an associative Hopf algebra was proposed in [6], where tensorial degrees
of freedom corresponding to the Lorentz generators were added to the position operator algebra.

This idea was then developed in a series of papers [7-8] using methods of realization of quantum phase spaces in terms of Heisenberg algebra
[5-9]; the algebra that included the tensorial generators was named extended Snyder algebra, to distinguish it from the standard realization
of the Snyder model in terms of
vectorial \dof only (called standard Snyder model in the following). Also generalizations including \kp deformations [8], and the
construction of an Heisenberg double for these algebras have been investigated [10].

Although this framework solves the mathematical problem related to the definition of a proper Hopf algebra, the physical interpretation
of the antisymmetric degrees of freedom is not obvious.

In this paper, we shall attempt to investigate a quantum mechanical model based on the Euclidean version of the extended Snyder model
and inspired by an analogous one introduced in the context of Moyal space, where the objects of noncommutativity were considered as antisymmetric
 operators [11].
In particular, we study the harmonic oscillator in this theory, with the aim of understanding in a simple case
the physical implications of the addition of the tensorial degrees of freedom, comparing the results with those obtained in [12,13] for
the standard Snyder model.

\section{2. The Snyder model}

We recall that the original $D$-dimensional Euclidean Snyder model is defined by the \cor
$$\eqalignno{&[\hx_i,\hx_j]=i\b^2M_\ij,\qquad[M_\ij,\hx_k]=i(\d_{ik}\hx_j-\d_{jk}\hx_i),&\cr
&[M_\ij, M_{kl}]=i(\d_{ik}M_{jl}-\d_{il}M_{jk}-\d_{jk}M_{il}+\d_{jl}M_{ik}),&(1)}$$
where latin indices run from 1 to $D$ and $\hx_i$ are position generators, $M_\ij$ rotation generators\footnote{$^1$}
{Of course, in the Euclidean case the Lorentz algebra is replaced by the algebra of rotations in $D$ dimensions.} and $\b$ is a real parameter,
that can be identified with the \nc Snyder parameter, which is usually assumed to be of the scale of the Planck length $L_p$.
For $\b=0$, the \cor (1) reduce to those of the standard rotation algebra acting on commutative coordinates.
One can then extend the model to phase space adding the momenta $\hp_i$ conjugated to $\hx_i$, satisfying
$$[\hp_i,\hp_j]=0,\qquad[\hx_i,\hp_j]=i\(\d_\ij+\b^2\hp_i\hp_j\),\quad
[M_\ij,\hp_k]=i(\d_{ik}\hp_j-\d_{jk}\hp_i).\eqno(2)$$
By choosing $M_\ij=x_ip_j-x_jp_i$ and finding suitable representations for $\hx_i$ and $\hp_i$, the algebra (1)-(2) can be realized in terms of
a canonical phase space of \coo $x_i$, $p_i$. In this way, it is possible to construct a coalgebra structure [5] and to define a star product.
However, since the algebra generated by the $\hx_i$ does not close, the resulting coproduct is not coassociative and the star product is not associative [5].

To remedy this, one can define an extended Snyder algebra, by promoting the $M_\ij$ to $D(D-1)/2$ noncommutative
tensorial degrees of freedom $\hx_\ij=-\hx_{ji}$ to be added to the $D$ position operators $\hx_i$ [6]. The total number of \dof is then
$D(D+1)/2$. The $D$-dimensional Euclidean extended Snyder algebra takes therefore the form
$$\eqalignno{&[\hx_i,\hx_j]=i\l\b^2\hx_\ij,\qquad[\hx_\ij,\hx_k]=i\l(\d_{ik}\hx_j-\d_{jk}\hx_i),&\cr
&\ \ [\hx_\ij, \hx_{kl}]=i\l(\d_{ik}\hx_{jl}-\d_{il}\hx_{jk}-\d_{jk}\hx_{il}+\d_{jl}\hx_{ik}),&(3)}$$
where we have introduced a deformation parameter $\l$, which in natural units is dimensionless.
Note that we assume that the coordinates $\hx_i$ have dimension of length, while the tensorial \coo $\hx_\ij$ are dimensionless, like the $M_\ij$,
although in this formalism the $\hx_\ij$ are no longer identified with the rotation generators.

Again, one may extend the algebra to phase space, by introducing the momenta $\hp_i$ and $\hp_\ij=-\hp_{ji}$, conjugated to the $\hx_i$ and $\hx_\ij$,
respectively.
This can be done in several inequivalent ways compatible with the Jacobi identities, that correspond to different realizations of the model\footnote
{$^2$}{At order $\l$ there exists a one-parameter family of realizations [7]. The realization (5) corresponds to the case $c_1=0$ of [7].} [7].
For the moment, we consider the so-called Weyl realization, for which at leading order in $\l$,
$$\eqalignno{&[\hp_i,\hp_j]=[\hp_\ij,\hp_{kl}]=[\hp_i,\hp_{jk}]=0,\qquad[\hx_i,\hp_{jk}]=i{\l\b^2\over2}(\d_{ik}\hp_j-\d_{jk}\hp_i),
\qquad[\hx_i,\hp_j]=i\(\d_\ij+{\l\over2}\hp_\ij\),&\cr
&[\hx_\ij,\hp_k]=i{\l\over2}(\d_{ik}\hp_j-\d_{jk}\hp_i),\qquad[\hx_\ij,\hp_{kl}]=i\(\d_{ik}\d_{jl}+{\l\over2}(\d_{ik}\hp_{jl}-\d_{il}\hp_{jk})-(k\lra l)\).&(4)}$$

However, we remark that a realization closer to (2) is given by what we may call classical realization, defined so that the \cor (2) hold
at order $\l^2$ (in particular, $[\hx_i,\hp_j]=i\(\d_\ij+\l^2\b^2\hp_i\hp_j\)$). In this case, the full set of \cor at order $\l$ is given by [8]
$$\eqalignno{&[\hp_i,\hp_j]=[\hp_\ij,\hp_{kl}]=[\hp_i,\hp_{jk}]=0,\qquad[\hx_i,\hp_{jk}]=i{\l\b^2\over2}(\d_{ik}\hp_j-\d_{jk}\hp_i),\qquad[\hx_i,\hp_j]=i\d_\ij,&\cr
&[\hx_\ij,\hp_k]=i\l(\d_{ik}\hp_j-\d_{jk}\hp_i),\qquad[\hx_\ij,\hp_{kl}]=i\(\d_{ik}\d_{jl}+{\l\over2}(\d_{ik}\hp_{jl}-\d_{il}\hp_{jk})-(k\lra l)\).&(5)}$$

Also for the \esa it is possible to construct realizations in terms of an extended Heisenberg algebra [7], obtained by adding tensorial
\dof $x_\ij=-x_{ji}$ to the standard Heisenberg algebra, as
$$\eqalignno{&[x_i,x_j]=[p_i,p_j]=[x_\ij,x_{kl}]=[p_\ij,p_{kl}]=0,&\cr
&[x_i,p_j]=i\d_\ij,\qquad [x_\ij,p_{kl}]=i(\d_{ik}\d_{jl}-\d_{il}\d_{jk}),&\cr
&[x_i,x_{jk}]=[x_i,p_{jk}]=[x_\ij,x_k]=[x_\ij,p_k]=0,&(6)}$$
where $p_i$ and $p_\ij$ are momenta canonically conjugate to $x_i$ and $x_\ij$ respectively, and $p_\ij=-p_{ji}$.

A simplification of the formalism can be obtained noticing that for $\b\ne0$ the algebra (3) is isomorphic to $so(D+1)$, so that it is convenient to define new variables [7]
$$\hx_i=\b\hx_{i,D+1},\qquad\hp_i={\hp_{i,D+1}\over\b},\eqno(7)$$
such that the algebra (3) takes the form
$$[\hx_\mn,\hx_{\rs}]=i\l(\d_{\m\r}\hx_{\n\s}-\d_{\m\s}\hx_{\n\r}-\d_{\n\r}\hx_{\m\s}-\d_{\n\s}\hx_{\m\r}),\eqno(8)$$
with Greek indices running from 1 to $N+1$.

The same can be done for the \eha (6), that becomes
$$[x_\mn,x_{\r\s}]=[p_\mn,p_{\r\s}]=0,\qquad[x_\mn,p_{\r\s}]=i(\d_{\m\r}\d_{\n\s}-\d_{\m\s}\d_{\n\r}),\eqno(9)$$

The algebra (3) can then be realized in terms of the \eha as a power series: in the Weyl realization one has at first order in $\l$,
$$\hat x_\mn=x_\mn+{\l\over 2}(x_{\m\a}p_{\n\a}-x_{\n\a}p_{\m\a}),\eqno(10)$$
while $\hp_\mn=p_\mn$. In terms of components
$$\eqalignno{\hx_i&=x_i+{\l\over2}\big(x_kp_{ik}-\b^2x_{ik}p_k\big),&\cr
\hx_\ij&=x_\ij+{\l\over2}\big(x_ip_j+x_{ik}p_{jk}-(i\leftrightarrow j)\big).&(11)}$$

It may be interesting to consider the symmetries of the extended Snyder algebra.
The algebra (3)-(4) is covariant under the action of the group $SO\({(D-1)D(D+1)(D+2)\over8}\)$ generated by
$L_{\mn,\r\s}=x_\mn p_\rs-x_\rs p_\mn$.
However, from a physical standpoint it is more interesting to consider its subgroup corresponding to rotations of
the $D$-dimensional space, with generators
$$M_\ij=x_ip_j-x_jp_i+x_{ik}p_{jk}-x_{jk}p_{ik}.\eqno(12)$$
acting as
$$[M_\ij,\hx_k]=i(\d_{ik}\hx_j-\d_{jk}\hx_i),\qquad [M_\ij,\hx_{jk}]=i(\d_{ik}\hx_{jl}-\d_{il}\hx_{jk}-\d_{jk}\hx_{il}+\d_{jl}\hx_{ik}).\eqno(13)$$

\section{3. The harmonic oscillator}

To test the dynamics of the model, we consider an isotropic harmonic oscillator.
We start by defining an Hamiltonian invariant under the extended Snyder symmetry, namely
$$H=\qu\sum_\mn\({\hp_\mn^2\over M}+M\o^2\hx_\mn^2\).\eqno(14)$$
where $M$ has dimension of length. Substituting the realization (10) we obtain at leading order in $\l$,
$$H=\qu\sum_\mn\[{p_\mn^2\over M}+M\o^2\Big(x_\mn^2+{\l^2\over2}x_{\m\r}p_{\n\r}(x_{\m\s}p_{\n\s}-x_{\n\s}p_{\m\s})\Big)\],\eqno(15)$$
Notice that terms of order $\l$ in the Hamiltonian vanish. The factor ${1\over4}$ is due to the fact that antisymmetric
\dof are counted twice.

To discuss the spectrum, it is useful to introduce creation and annihilation operators,
$$a_\mn=\sqrt{M\o\over2}\(x_\mn+i\,{p_\mn\over M\o}\), \qquad\ad_\mn=\sqrt{M\o\over2}\(x_\mn-i\,{p_\mn\over M\o}\),\eqno(16)$$
satisfying $a_\mn=-a_\nm$, $\ad_\mn=-\ad_\nm$, and
$$[a_\mn,\ad_{\r\s}]=\d_{\m\r}\d_{\n\s}-\d_{\n\r}\d_{\m\s}.\eqno(17)$$
It is then convenient to split the Hamiltonian in a free part $H_0$ and an interaction term $V$, with $H=H_0+\l^2V$, such that
$$H_0=\qu\sum_\mn\({p_\mn^2\over M}+M\o^2x_\mn^2\),\qquad V={M\o^2\over8}\Big(x_{\m\r}p_{\n\r}(x_{\m\s}p_{\n\s}-x_{\n\s}p_{\m\s})\Big).\eqno(18)$$
One has then
$$H_0={\o\over4}\sum_\mn(a_\mn\ad_\mn+\ad_{\m\n}a_{\m\n})={\o\over2}\(\sum_{\m\ne\n} N_\mn+{D(D+1)\over2}\),\eqno(19)$$
with $N_\mn=\ad_\mn a_\mn$, so that $N_\mn=N_\nm$, $N_{\m\m}=0$, and
$$V=-{M\o^2\over16}\sum_{\m\n}\sum_{\r\s}(a_{\m\r}\ad_{\n\r}-\ad_{\m\r}a_{\n\r})(\ad_{\m\s}a_{\n\s}-\ad_{\n\s}a_{\m\s}),\eqno(20)$$
In (20) we have retained only the terms that can contribute to the leading-order corrections to the energy.
After some manipulations, the interaction term reduces to
$$V={M\o^2\over8}\(\sum_{\m\ne\n}\sum_\r N_{\m\r}N_{\n\r}+(D-1)\sum_{\m\ne\n}N_{\mn}\).\eqno(21)$$

The free part of the Hamiltonian describes an harmonic oscillator in canonical extended spacetime, which we will call canonical extended oscillator.
Its energy spectrum is, as could be expected, that of an harmonic oscillator with $D(D+1)/2$ degrees of freedom.

In fact, defining the occupation numbers $n_\mn$ such that $N_\mn\,|\dots,n_\mn,\dots>\ =n_\mn\,|\dots,n_\mn,\dots>$,
the energy corresponding to a set of occupation numbers $\{n_\mn\}$ is at order 0 (\ie for the canonical extended oscillator)
$$E_{\{n_\mn\}}={\o\over2}\(\sum_{\m\ne\n} n_\mn+{D(D+1)\over2}\),\eqno(22)$$
while the leading-order corrections due to the Snyder structure are given by
$$\D E_{\{n_\mn\}}=\ <\{n_\mn\}|\l^2V|\{n_\mn\}>\ ={\l^2\b^2m\o^2\over8}\(\sum_{\m\ne\n}\sum_\r n_{\m\r}n_{\n\r}+(D-1)\sum_{\m\ne\n}n_\mn\).\eqno(23)$$
Hence, while the canonical extended oscillator has a standard spectrum depending only on the quantum number $n=\sum_{\m<\n}n_\mn$,
the Snyder extended oscillator has eigenvalues that depend on combinations of all the quantum numbers $n_\mn$.
The order of magnitude of the corrections to the energy spectrum (for $\l\approx1$) is the same as in the standard Snyder oscillator [12,13].

\section{4. Noncovariant formalism}

To better understand the physics, it is however useful to separate vector and tensor degrees of freedom, studying the model
from a $D$-dimensional point of view. Then the Hamiltonian can be written as
$$H=\ha\sum_i\({\hp_i^2\over m}+m\o^2\hx_i^2\)+\qu\sum_\ij\({\hp_\ij^2\over M}+M\o^2\hx_\ij^2\),\eqno(24)$$
where we have identified $m=M\b^{-2}$ with the mass of the vectorial \dof (i.e.~standard position coordinates)\footnote{$^3$} {Note that in the limit
$\b\to0$, $m$ diverges if $M$ is finite. However, as we shall see, the energy spectrum is regular and goes to the canonical one for $\b\to0$.}.
Then the free Hamiltonian reads
$$H_0=\ha\sum_i\({p_i^2\over m}+m\o^2x_i^2\)+\qu\sum_\ij\({p_\ij^2\over M}+M\o^2x_\ij^2\),\eqno(25)$$
and the interaction term becomes
$$\eqalignno{V={M\o^2\over8}\sum_\ij&\Big(x_ip_j(x_ip_j-x_jp_i)+x_{ik}p_{jk}(x_{il}p_{jl}-x_{jl}p_{il})+2x_ip_j(x_{ik}p_{jk}-x_{jk}p_{ik})&\cr
&+\b^{-2}x_jp_{ij}x_kp_{ik}+\b^2x_{ij}p_jx_{ik}p_k-x_ip_jp_{jk}x_{ik}-p_jx_ix_{ik}p_{jk}\Big).&(26)}$$



Defining, for $\b\ne0$,
$$a_i=\sqrt{M\o\over2}\({x_{4i}\over\b}+i\,{\b p_{4i}\over M\o}\), \qquad\ad_i=\sqrt{M\o\over2}\({x_{4i}\over\b}-i\,{\b p_{4i}\over M\o}\),\eqno(27)$$
the free Hamiltonian takes the form
$$H_0=\o\(\sum_i\ad_i a_i+{D\over2}+\ha\sum_{i\ne j}\ad_\ij a_\ij+{D(D-1)\over4}\),\eqno(28)$$
with spectrum
$$E_{\{n_i,n_\ij\}}=\o\(\sum_i n_i+\ha\sum_{i\ne j}n_\ij+{D(D+1)\over4}\),\eqno(29)$$
where $n_i$, $n_\ij$ are the occupation numbers for the vector and tensor degrees of freedom.
The leading-order corrections due to the Snyder structure arising from (24) are instead
$$\D E_{\{n_i,n_\ij\}}={\l^2\b^2m\o^2\over8}\[\sum_{i\ne j}n_{ik}n_{jk}+\sum_{i\ne j}n_in_j+2\sum_{ik}n_kn_{ik}+(D-1)\(\sum_{i\ne j}n_\ij+2\sum_i n_i\)\].\eqno(30)$$
These results are in accordance with (23) and can be compared with the spectrum of the standard Snyder oscillator [12,13], for which
$E_n\sim\o\(\sum_in_i+{D\over2}\)+o(\b^2m\o)$. It turns out that the vacuum energy is different in the two cases, while the leading order correction, although
different, are of the same order of magnitude. Notice also that the higher-order corrections to the vacuum energy (which vanish in our calculations)
depend on the specific operator ordering chosen.
\medskip

However, in this context, it seems more reasonable to choose a different Hamiltonian, invariant only under the $D$-dimensional rotation
group. One can still adopt the same  expression for the kinetic term, but assume different frequencies $\o$ and $\O$ for the vector and tensor \dof
in the interaction term, namely
$$H=\sum_i\({\hp_i^2\over2m}+{m\o^2\over2}\hx_i^2\)+\sum_\ij\({\hp_\ij^2\over4M}+{M\O^2\over4}\hx_\ij^2\).\eqno(31)$$
Using the realization (11) for $\hx_i$ and $\hx_\ij$, the spectrum of the free Hamiltonian results
$$E_{\{n_i,n_\ij\}}=\o\(\sum_i n_i+{D\over2}\)+{\O\over2}\(\sum_{i\ne j}n_\ij+{D(D-1)\over2}\),\eqno(32)$$
while
$$\eqalignno{V&={M\o^2\over8}\sum_{ijk}\Big(\b^2x_{ij}p_jx_{ik}p_k+\b^{-2}x_jp_{ij}x_kp_{ik}-x_ip_jp_{ik}x_{jk}-p_ix_jx_{ik}p_{jk}\Big)&\cr
&+{M\O^2\over4}\bigg(\sum_\ij x_ip_j(x_ip_j-x_jp_i)+\sum_{ijkh}x_{ik}p_{jk}(x_{ih}p_{jh}-x_{jh}p_{ih})+2\sum_{ijk}x_ip_j(x_{ik}p_{jk}-x_{jk}p_{ik})\bigg),&(33)}$$
and therefore
$$\eqalignno{\D E_{\{n_i,n_\ij\}}=&\ {\l^2\b^2m\over8}\bigg[\O^2\Big(\sum_{i\ne j}n_{ik}n_{jk}+\sum_{i\ne j}n_in_j\Big)+2\o^2\sum_{i\ne j}n_in_\ij+(D-1)(\O^2+\o^2)\sum_i n_i&\cr
&+\Big((D-2)\O^2+\o^2\Big)\sum_\ij n_\ij\bigg].&(34)}$$

It is reasonable to assume $\O\gg\o$.
In this case, one can make the approximation that the tensorial degrees of freedom are in the ground state, and then
$$E_{\{n_i,0\}}\sim\o\sum_i n_i+\({D\o\over2}+{D(D-1)\O\over4}\)+{\l^2\b^2m\over8}\(2(D-1)(\o^2+\O^2)\sum_i n_i+\O^2\sum_{i\ne j}n_in_j\).\eqno(35)$$
It is evident that the vacuum energy and the order-$\b^2$ corrections are greatly increased \wrt the standard Snyder oscillator,
even if the order of magnitude depends on the ratio $\O/\o$.

\medskip
As we have mentioned, the energy spectrum depends on the realization [13].
For example, let us consider the classical realization, with \cor (3), (5).
This can be obtained by setting [7]
$$\eqalignno{\hx_i&=x_i-{\l\b^2\over2}x_{ik}p_k,&\cr
\hx_\ij&=x_\ij+{\l\over2}\Big(2x_ip_j+x_{ik}p_{jk}-(i\leftrightarrow j)\Big).&(36)}$$
In this case, the zeroth-order energy (32) is of course unchanged, while the correction terms give rise to a different potential, namely
$$\eqalignno{V=&\ {M\o^2\over8}\sum_{ijk}\b^2x_{ij}p_jx_{ik}p_k+{M\O^2\over8}\Bigg(4\sum_\ij x_ip_j(x_ip_j-x_jp_i)&\cr
&+\sum_{ijkh}x_{ik}p_{jk}(x_{ih}p_{jh}-x_{jh}p_{ih})+4\sum_{ijk}x_ip_j(x_{ik}p_{jk}-x_{jk}p_{ik})\Bigg).&(37)}$$
A calculation analogous to the previous one gives for the leading order corrections to the energy
$$\eqalignno{\D E_{\{n_i,n_\ij\}}=&\ {\l^2\b^2m\over8}\bigg[\O^2\Big(\sum_{i\ne j}n_{ik}n_{jk}+4\sum_{i\ne j}n_in_j\Big)+\o^2\Big(\sum_{i\ne j}n_in_\ij+{D(D-1)\over4}\Big)&\cr
&+(D-1)\Big(4\O^2+{\o^2\over2}\Big)\sum_i n_i+\Big((D-2)\O^2+{\o^2\over2}\Big)\sum_\ij n_\ij\bigg].&(38)}$$
Hence, although the structure of the corrections is identical to that obtained for the Weyl realization, the numerical coefficients are rather different.
This is a typical feature of noncommutative models, where, for a given Hamiltonian, different realizations lead to nonequivalent physical models [13,14].

\section{5. Conclusions}
The extended Snyder model includes tensorial degrees of freedom in addition to the standard vectorial ones, allowing a more satisfying definition of its associated
Hopf algebra.
We have considered an harmonic oscillator in the context of this model, and calculated its energy spectrum.
It results that the corrections to the spectrum are of the same order $\b^2m\o$ as in the standard Snyder model [12,13]. However, if one allows for different frequencies
to be associated to vectorial and tensorial degrees of freedom, the magnitude of the corrections can increase much, depending on the ratio of the two frequencies.

We have assigned to the tensorial degrees of freedom a null physical dimension in natural units, as for the angular momentum.
However, we should mention the possibility of assigning them a noncanonical dimension of length, so that it coincides with the one of the vectorial degrees of freedom.
In this case, one can associate a mass $m$ to the the tensorial variables identical to that of the vectors. The conclusions about the harmonic oscillator are unaffected,
since its properties do not depend on the mass, but the properties of more complex models could depend on this choice. For example, if the tensorial variables very weakly
interact with the vectors, they constitute a huge hidden mass whose interaction with ordinary matter can be hardly detectable, and could allow the construction of models
for dark matter.

As discussed in the appendix, similar results are obtained in the case of a Moyal model in which the \oa is promoted to a dynamical variable,
as proposed in [11]. Differences arise only in the details of the leading-order corrections to the energy. This fact suggests us the conjecture that all noncommutative
models which include antisymmetric dynamical variables lead to the same structure when applied to the harmonic oscillator problem.

Another effect that was pointed out in [11] is the fact that the uncertainty relations can be modified. This happens also in our case, but is realization dependent.
For example, it is clear from (4) that the uncertainty relations for $\D x_i\D p_j$ in the Weyl realization depend
on the expectation values of the tensorial degrees of freedom, while in the classical realization (5) they coincide with those of the standard Snyder model.
Also, modifications to the Casimir force between conducting plates could be evaluated on the lines of the calculations performed in [15] for the standard case.

Our investigation can easily be extended to a relativistic setting  on the lines of [16]. More interesting would be to define a quantum field theory on
the extended Snyder background, that could solve some of the problems found in the standard theory [17]. Of course, the introduction of the new variables would greatly
modify the formalism.
\bigbreak

\acknowledgments
S. Mignemi acknowledges contribution from GNFM and COST action CA18108.

\beginref
\ref [1] H.S. Snyder, \PR{71}, 38 (1947).
\ref [2] S. Doplicher, K. Fredenhagen and J. E. Roberts, \PL{B331}, 39 (1994).
\ref [3] J. Lukierski, H. Ruegg, A. Novicki and V.N. Tolstoi, \PL{B264}, 331 (1991).
\ref [4] S. Majid,  {\it Foundations of quantum group theory}, Cambridge University Press 1995. ??
\ref [5] M.V. Battisti and S. Meljanac, \PR{D79}, 067505 (2009);
 M.V. Battisti and S. Meljanac, \PR{D82}, 024028 (2010).
\ref [6] F. Girelli and E. Livine, \JHEP{1103}, 132 (2011).
\ref [7] S. Meljanac and S. Mignemi, \PR{D102}, 126011 (2020).
\ref [8] S. Meljanac and S. Mignemi, \PL{B814}, 136117 (2021);
S. Meljanac and S. Mignemi, \PR{D104}, 086006 (2021).
\ref [9] S. Meljanac, D. Meljanac, A. Samsarov and M. Stoji\'c, \PR{D83}, 065009 (2011);
T. Juric, S. Meljanac, D. Pikuti\'c and R. \v Strajn, \JHEP{07}, 055 (2015).
S. Meljanac and R. \v Strajn, SIGMA {\bf 18}, 022 (2022)
\ref [10] S. Meljanac and A. Pachol, Symmetry {\bf 13}, 1055 (2021).
\ref [11] R. Amorim, \PRL{101}, 081602 (2008).
\ref [12] L.N. Chang, D. Minic, N. Okamura and T. Takeuchi, \PR{D65}, 125027 (2002);
S. Mignemi, \PR{D84}, 025021  (2011).
\ref [13] G. Gubitosi and S. Mignemi, Universe {\bf 8}, 108 (2022).
\ref [14] G. Amelino-Camelia, S. Bianco and G. Rosati, \PR{D101}, 026018 (2020).
\ref [15] S. Franchino-Vi\~nas and S. Mignemi, \NP{B959}, 115152 (2020).
\ref [16] R. Amorim, E.M.C. Abreu and W.G. Ramirez, \PR{D81}, 105005 (2010).
\ref [17] S. Meljanac, S. Mignemi, J. Trampeti\'c and J. You, \PR{D96}, 045021 (2017);
S. Meljanac, S. Mignemi, J. Trampeti\'c and J. You, \PR{D97}, 055041 (2018);
S. Franchino-Vi\~nas and S. Mignemi, \PR{D98}, 065010 (2018).
\endref
\section{Appendix}
In this appendix we compare our results with those arising from a Moyal oscillator with dynamical noncommutativity [11]. A similar calculation has been
performed in [11], but the author employed a different approach, in particular choosing a deformed Hamiltonian, such that the energy spectrum maintains
its canonical form.

We do not report here the details of the computation, since they are analogous to those performed in the Snyder case.
The commutation relations of the Moyal space are [2]
$$[\hx_i,\hx_j]=i\l\,\h_\ij,\qquad[\hp_i,\hp_j]=0,\qquad[\hx_i,\hp_j]=i\d_\ij,\eqno(A.1)$$
where the \oa $\h_\ij$ is a constant antisymmetric tensor of dimension inverse length square and $\l$ a dimensionless deformation parameter.
In [11] it was proposed to promote $\h_\ij$ to an independent dynamical variable $\hx_\ij$ with conjugate momentum $\hp_\ij$, in order to maintain
rotational covariance.

One has then,
$$\eqalignno{&[\hx_i,\hx_j]=i\l\,\hx_\ij,\qquad[\hp_i,\hp_j]=0,\qquad[\hx_i,\hp_j]=i\d_\ij,&\cr
[\hx_\ij,\hp_{kl}]&=i(\d_{ik}\d_{jl}-\d_{il}\d_{jk}),\qquad[\hx_i,\hp_{jk}]=-i{\l\over2}(\d_{ik}\hp_l-\d_{il}\hp_k),&\cr
&[\hx_i,\hx_{jk}]=[\hx_\ij,\hx_{kl}]=[\hp_i,\hx_{jk}]=[\hp_i,\hp_{jk}]=0.&(A.2)}$$
The \cor (A.2) are similar to those of the extended Snyder model and can be obtained analogously in terms of the \eha (6), defining
$$\hx_i=x_i-{\l\over2}x_\ij x_ip_j,\qquad\hp_i=p_i,\qquad\hx_\ij=x_\ij,\qquad \hp_\ij=p_\ij.\eqno(A.3)$$
Contrary to ref.~[11] we choose the standard Hamiltonian (31) for the extended harmonic oscillator.
This will give rise to corrections to the energy spectrum of the canonical extended oscillator.

In fact, substituting (A.3) in (31) we obtain an effective Hamiltonian in terms of canonical operators $x_i$, $p_i$, $x_\ij$, $p_\ij$,
$$H=\sum_i\({p_i^2\over2m}+{m\o^2\over2}\Big(x_i^2-\l\,x_\ij x_ip_j+{\l^2\over4}x_\ij p_jx_{ik}p_k\Big)\)
+\sum_\ij\({p_\ij^2\over4M}+{M\O^2x_\ij^2\over4}\).\eqno(A.4)$$
As before, this can be separated in a free part $H_0$ (25) and an interaction part. The free part has of course the same spectrum as in the extended
Snyder model.
The leading-order corrections to the energy come instead from the term
$$V={\l^2m\o^2\over8}\ x_\ij p_jx_{ik}p_k,\eqno(A.5)$$
which also appears in (37).

We then go through the same passages as in the Snyder case, obtaining
$$\eqalignno{E_{\{n_i,n_\ij\}}&=\o\(\sum_i n_i+{D\over2}\)+{\O\over2}\(\sum_{i\ne j}n_\ij+{D(D-1)\over2}\)&\cr
&+{\l^2m\o^2\over8}\(\sum_{i\ne j}n_in_\ij+{D-1\over2}\sum_i n_i+\ha\sum_\ij n_\ij+{D(D-1)\over4}\).&(A.6)}$$
Although the details are of course different, this result is qualitatively similar to the one obtained in the Snyder case.
It is likely that analogous results hold for any noncommutative model containing antisymmetric degrees of freedom.

\end